# Interaction-Free Effects Between Distant Atoms

Yakir Aharonov[1,2,4], Eliahu Cohen[3,4], Avshalom C. Elitzur[4], Lee Smolin[5]

[1]School of Physics and Astronomy, Tel Aviv University, Tel-Aviv 6997801, Israel
yakir@post.tau.ac.il

[2]Schmid College of Science, Chapman University, Orange, CA 92866, USA

[3]H.H. Wills Physics Laboratory, University of Bristol, Tyndall Avenue, Bristol, BS8 1TL, U.K
eliahu.cohen@bristol.ac.uk

[4]Iyar, The Israeli Institute for Advanced Research, POB 651, Zichron Ya'akov 3095303, Israel
avshalom@iyar.org.il

[5]Perimeter Institute for Theoretical Physics, 31 Caroline Street North, Waterloo, Ontario N2J 2Y5, Canada
lsmolin@perimeterinstitute.ca

*A Gedanken experiment is presented where an excited and a ground-state atom are positioned such that, within the former's half-life time, they exchange a photon with 50% probability. A measurement of their energy state will therefore indicate in 50% of the cases that no photon was exchanged. Yet other measurements would reveal that, by the mere possibility of exchange, the two atoms have become entangled. Consequently, the "no exchange" result, apparently precluding entanglement, is non-locally established between the atoms by this very entanglement. This quantum-mechanical version of the ancient Liar Paradox can be realized with already existing transmission schemes, with the addition of Bell's theorem applied to the no-exchange cases. Under appropriate probabilities, the initially-excited atom, still excited, can be entangled with additional atoms time and again, or alternatively, exert multipartite nonlocal correlations in an interaction free manner. When densely repeated several times, this result also gives rise to the Quantum Zeno effect, again exerted between distant atoms without photon exchange. We discuss these experiments as variants of Interaction-Free-Measurement, now generalized for both spatial and temporal uncertainties. We next employ weak measurements for elucidating the paradox. Interpretational issues are discussed in the conclusion, and a resolution is offered within the Two-State Vector Formalism and its new Heisenberg framework.*

Wave-particle duality, nonlocality, and the measurement problem are often considered as quantum mechanics' most fundamental paradoxes. This triad, however, does not exhaust the theory's uniqueness. No less paradoxical is the causal efficacy of counterfactual quantum events. Consider, *e.g.*, Interaction-Free Measurement (IFM) [1]: A particle *may* hit a detector but eventually *does not*, yet the former's momentum *does* change, just because it *could* have hit the latter. Several related effects, such as Hardy's paradox [2], intensify this quantum oddity.

A simple asymmetric interaction between two particles, named Quantum Oblivion [3-5], has recently revealed the mechanism underlying all these "could" phenomena. After this interaction, one of two the particles undergoes momentum change while the other remains unaffected. A more detailed analysis reveals that, during a very brief interval, entanglement has been formed and then gone away. This, then, is what happens in IFM:

A brief moment before the measurement is finalized, partial entanglement *is* formed between them, immediately to be reversed. Consequently, the measured particle undergoes momentum change while no matching change can be observed in the detector. Under such finer time-resolution, many varieties of quantum measurement, *e.g.*, the AB effect [6], the quantum Zeno effect [7] and quantum erasure [8], similarly turn out to stem from Quantum Oblivion [3].

Such is the *Gedanken experiment* proposed below. A photon, which *could* have been emitted, entangles two distant atoms, making them EPR entangled, yet a measurement of this photon's whereabouts may reveal the petty fact that it still resides where it has initially been.

While of our basic thought-experiment's main setup is widely used, it is a particular possible outcome of it, hitherto unnoticed, which reveals the quantum surprise and its consequences discussed below.

We begin our analysis with a well-known setup [9], which nevertheless reveals fundamental and somewhat surprising effects. We then generalize and scrutinize it using strong/weak measurements for understanding the implications of the apparent paradoxical behavior.

This paper's outline is as follows. Sec. 1 presents the proposed Gedanken experiment and Sec. 2 analyzes the predicted results. In 3-4 it is shown that the no-exchange outcome may repeat itself several times, a phenomenon that 5shows to indicate the involvement of the Quantum Zeno effect, and moreover that the latter effect itself is analogous to our scheme. In 6 we add weak measurements [10-12] to take advantage of non-commuting measurements that would otherwise remain merely counterfactual, yet can be weakly employed alongside with the actual strong ones. Sec.7 discusses the paradox in the broader context of several related quantum effects and reveals their underlying affinity. 8 is an extended discussion on the results and their significance.

1. **Paying for Entanglement with a Photon that is Not Emitted**

For the present Gedanken level, idealized settings will suffice, ignoring several technical issues. The latter are dealt extensively by [9] and others, who use this scheme for practical purposes, thereby offering a realizable setting for the present foundational issues.

Place an excited atom *A* at $t = 0$ inside a long reflecting cavity, such that, upon decaying, it emits a photon straight along the cavity's opening direction (Fig. 1). Emission will occur under the time-energy uncertainty

$$|\psi(t)\rangle_A = 2^{-t/2\tau}|e\rangle_A + \sqrt{1-2^{-t/\tau}}|g\rangle_A, \qquad \tau\Delta E \geq \frac{\hbar}{2}, \tag{1}$$

where $\tau$ is the atom's half-life time and $\Delta E$ the difference between its two energy levels. Next place another atom *B*, of the same element but in a ground state, within an identical cavity, located at distance *d* from *A* and oppositely facing it (Fig. 1). Wait for the excited atom's half-life time $\tau \ll d/c$ (to prevent multiple emissions and absorptions). By Eq. 1, the atom has emitted the photon with *P*=1/2. Now close *A*'s cavity door and wait till $\tau + d/c$ to close *B*'s cavity door as well. By virtue of the possibility of photon exchange, the two atoms' states have become entangled:

$$|\psi(t)\rangle_{AB} = \sqrt{1-\varepsilon^2}\left(2^{-(t-t_i)/2\tau}|e\rangle_A|g\rangle_B - \sqrt{1-2^{-(t-t_i)/\tau}}|g\rangle_A|e\rangle_B\right) + \varepsilon|g\rangle_A|g\rangle_B|\gamma\rangle. \tag{2}$$

The $\varepsilon^2$ constant accounts for the chance of finding the two atoms in their ground states while the photon $\gamma$ is still traveling along the route connecting them, which we rather avoid, although the proposed paradox would in fact persist in a weaker form even if $\varepsilon > 0$. But for simplicity we shall assume $\varepsilon$ is strictly zero by excluding all cases where a photon was caught on its way from *A* to *B*. The relative phase of $\pi$ was chosen for making the resulting state a singlet.

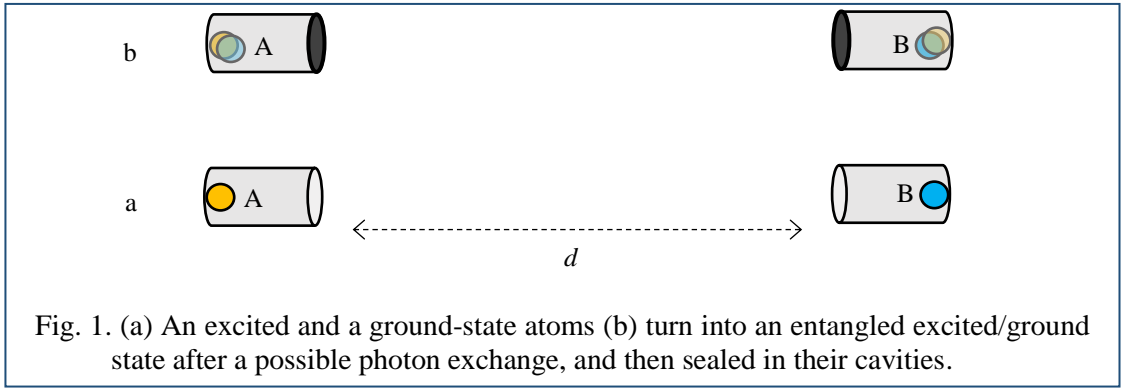

Fig. 1. (a) An excited and a ground-state atoms (b) turn into an entangled excited/ground state after a possible photon exchange, and then sealed in their cavities.

Proceed to prepare many such entangled pairs. For each pair, give atoms *A* and *B* to Alice and Bob, respectively, for EPR-Bell measurement (better rush within interval $\tau$ after the pair's preparation, to avoid re-emission in case *B* has absorbed the photon).

To prove Bell Inequality violations, thereby showing that the atoms' correlations are created nonlocally, Alice/Bob must randomly choose every time one out of three variables to measure on atom *A/B* of each pair. The first variable, naturally, is whether the atom is excited or ground. Two more variables need to be also available for choice, variables that maintain uncertainty relations with the first. The magnetic dipole moment $\vec{\mu}$ offers two such suitable variables. Upon applying the proper magnetic field on the atom and performing a projective measurement along the $\hat{\xi}$ direction, its magnetic moment's outcomes equal $\mu_\xi \equiv \frac{\vec{\mu} \cdot \hat{\xi}}{|\vec{\mu}|} = \pm 1$. These, in turn, correspond to linear combinations of the excited and ground states: $\cos\alpha |e\rangle + \sin\alpha |g\rangle$. Thus a measurement of $\mu_z$, i.e., in the $\alpha = 0$ direction, is essentially an energy measurement, resulting in $|e\rangle$ with either an eigenvalue $\mu_z = +1$ or $|g\rangle$ with an eigenvalue $\mu_z = -1$. A measurement of $\mu_\zeta$, for instance, where $\hat{\zeta} = \frac{\hat{x}+\hat{y}}{\sqrt{2}}$ corresponds to $\alpha = \pi/4$, can discern between the $\frac{1}{\sqrt{2}}(|e\rangle + |g\rangle)$ and $\frac{1}{\sqrt{2}}(|e\rangle - |g\rangle)$ states. Conversely, measurement in the *E* or $\mu_z$ bases corresponds to measurement in the $\alpha = 0$ direction. Measurements along other directions refer to rotations of the magnetic field.

This way, precisely like the three customary spin/polarization directions measured in ordinary EPR-Bell experiments, we have the three measurement-bases *E*, $\mu_{\zeta_1}$ and $\mu_{\zeta_2}$, with the analogous Bell correlations between them:

$$C(\alpha_1, \alpha_2) = -\cos(2(\alpha_2 - \alpha_1)), \tag{3}$$

where $\alpha_1$ and $\alpha_2$ are the angles chosen by Alice and Bob in the original EPR-Bell version, translated in our version to the above three choices.

## 2. Results and Interpretation

Having registered many such pairs of measurement outcomes, Alice and Bob now compare them similarly to a Bell test.

As the measurements' choices have been taken randomly, each of the partners has chosen to measure either *E*, $\mu_{\zeta_1}$ or $\mu_{\zeta_2}$, equally in ~1/3 of the cases. Consider then the case where Alice measures $\alpha_2 = \pi/3$, obtaining, say, +1. She can then infer the following about the other, remote atom:

1. If Bob measures $\alpha_2 = \pi/3$, he gets -1 in 100% of the cases.

2. If Bob measures $\alpha_1 = \pi/6$, he gets -1 in 75% of the cases.

3. If Bob measures $E$, he gets $|g\rangle_B$ in 25% of the cases.

And similarly for the second magnetic moment $\alpha_1 = \pi/6$.

These correlations are

a. Nonlocal: The dependence on the *relative* angle in Eq. (3) means: Each outcome obtained by Alice/Bob is quantumly correlated with the *random* outcome -1/+1 obtained by Bob/Alice plus their *deliberate* choice of the variables E/$\mu_{\varsigma_1}/\mu_{\varsigma_2}$ to which this outcome pertains.

b. Lorentz covariant: Each party's choice can equally be the partial "cause" or "effect" of the other's outcome, depending on the reference-frame.

A quantum paradox therefore ensues when Alice/Bob measures *E*, revealing whether the photon has been emitted. In half of these cases (total 1/6 of all measurements), *the initially excited/ground atom turns out to be still excited/ground*. Yet all the predictions derived from Bell's inequality hold for this case just as well. For example, Alice, having obtained $|e\rangle_A$, which indicates that her atom has *never* emitted its photon, is nevertheless informed by this outcome that:

4. If Bob measures *E*, he gets $|g\rangle_B$ (affirming that his atom has never absorbed the never-emitted photon) in 100% of the cases.

5. If Bob measures $\alpha_1 = \pi/3$, he gets +1 in 75% of the cases.

6. If Bob measures $\alpha_2 = \pi/6$, he gets +1 in 25% of the cases.

The above counterfactuals (1) and (4), obliging 100% correlations for the two other variables which were not prepared in advance, prove that the "excited"/"emitted" outcome is an equally nonlocal effect.[1]

Let us stress that any other attempt to exclude this *E-E* group of outcomes from Bell-inequality's jurisdiction in this case is as arbitrary as excluding *any* same-variable group from the standard Bell setting. It would be absurd, for example, to dismiss all *x-x* cases, arguing that their correlations could have emerged locally as well. In the present case such an *E-E* exclusion has no rationale other than wishing to escape the paradox. This objection becomes even more compelling by the above Lorentz covariance (b): It is equally Bob's choice that, by Bell's proof, can be interpreted as having effected Alice's "no emission" outcome.

---

[1] In other words, stating that the atom "has *remained* excited/ground" is somewhat misleading. Rather, it has been initially *prepared* excited/ground, then became *superposed*, and then sometimes *returned* to the original state, which can occur only via nonlocal correlation with the other atom.

No less paradoxical is the case when Alice finds that her atom *has* emitted its photon. By (b), Bob's choice between $E$, $\mu_{\varsigma_1}$ and $\mu_{\varsigma_2}$ is supposed to affect this emission of Alice's photon, which, to enable the entanglement facilitating this nonlocal effect, must have occurred earlier!

Let us summarize. *The indication of atom A/B that it has never emitted/absorbed a photon, which may naively suggest that it could not be entangled with B/A, is the result of this very A-B entanglement.* The classical liar paradox stemming from Epimenides' claim that "all Cretans are liars" is not necessarily absurd when stated by a quantum-mechanical Cretan (see also [13]). In Secs. 7-8 we would see that this kind of naïve reasoning is flawed.

The paradox is inherently quantum, based on the creation and validation of quantum entanglement, thus with no classical analogue. By now we got used to the strong nonlocal correlations enabled by entangled states, but here including the specific preparation method and maintain the information as to which atom was initially excited brings about a quantum paradox. The two atoms' quantum state is the one which, in terms of the PBR theorem, undergoes a change that is strictly ontic, although subtle, rather than merely an epistemic one concerning the observer's information about it. This straightforwardly follows from the ontic reading of the quantum state (i.e. from treating it as corresponding to physically real object) suggested by the PBR theorem [14].

### 3. Getting Away Further with Non-Payments

Having managed to form entanglement with a distant atom, yet with the photon you had to pay with still being with you, why not proceed to "double sting"? Once your measurement indicates that your atom *A* has emerged from the entanglement with *B* still excited, then simply rush to direct its cavity, like a torch, towards another oppositely-facing cavity with another ground-state atom *B'*, in order to create a new EPR pair. Your success probability for such double luck is, *a-priori*, 1/4, but once you *did* get away with the first atom, the probability goes up to 1/2 again.

Wish to push your luck further with *B″* and so on? Again, the *a-priori* probability for success goes down to $P = (1/2)^N$, but if you were lucky *N*-1 times, success chances for the $N^{th}$ are again 1/2. All you can lose is the single photon that, had you been playing fair, you would have given in the first time anyway.

To summarize, adding the "no emission" option to the scheme of Cirac *et al.*'s [9] protocol enables entangling the excited atom with several more partners. This holds for the next section as well.

### 4. Simultaneous Entangling of Multiple Atoms Through the Unemitted Photon

A more striking outcome emerges when your excited atom *A* is surrounded by a sphere with radius *d* of ground atoms $B_1, B_2, \ldots, B_{N-1}$. Is it possible to affect all of them *at once*? For this purpose, let *A* be out of its cavity to enable it to emit its photon to all directions, and wait. At $t = \tau'$, being the time in which the atom has become ground with probability *(N-1)/ N*, all *N* atoms become entangled

$$|\Psi_N\rangle = |W_N\rangle, \tag{4}$$

where $|W_N\rangle$ is the obviously nonlocal *N*-partite *W*-state [15]

$$|W_N\rangle = \frac{1}{\sqrt{N}} \left( |e\rangle|g\rangle|g\rangle \cdots |g\rangle + |g\rangle|e\rangle|g\rangle \cdots |g\rangle + \cdots + |g\rangle|g\rangle \cdots |g\rangle|e\rangle \right), \tag{5}$$

known to be important for quantum information applications, since it is robust against particle loss [16].

Now in *1/N* of the cases Alice, measuring her atom's energy, will find it still excited. But what about all the remote Bobs? If they measure their atoms' energies, they will always find them in *ground* state, but following a similar logic to that of Sec. 2, Alice and Bobs can verify nonlocal correlations between their atoms. For instance, they can empirically detect a violation of Bell inequality (see [17] for the case of *N*=3).

### 5. Zeno Collaborates with Epimenides

Despite its inherent indeterminism, quantum mechanics can also steer its randomness towards a desired direction. The Quantum Zeno effect [7], for example, has been employed to increase IFM's efficiency from $P = 1/2$ to $P \to 1$ [18]. Not only is this improvement equally applicable for our Liar paradox, but the Zeno effect itself turns out to be inherently embedded in our setting, in several interesting ways.

Recall first that the above nonlocal effects are Lorentz invariant, meaning that we can equally explain Bob's measurement outcome as the *cause* rather than the effect of Alice's measurement. Applied to the multiple entanglement scenarios described in Secs. 3-4 above, it is Bob's measurements that "rejuvenate" Alice's atom, enabling it to emerge excited time and again. While this effect is basically random, it can be manipulated into a systematic delay of atom $A$'s decay, which is the Quantum Zeno effect. All Bob should do is to perform a dense enough set of projective energy measurements on the potentially incoming photon, thereby preventing Alice's atom from becoming ground.

This is described in greater detail elsewhere [19]. In the present context it should be pointed out that many (if not all) standard Quantum Zeno demonstrations published so far turn out, in retrospect, to be inadvertent Quantum Liar experiments. After all, all Zeno experiments involve *non-clicking* of a photon detector directed to the measured atom. It is only the verification method, namely Bell's theorem, which is added in the present case. "By performing the null measurements frequently or continuously, one can freeze the spin dynamics. This is a kind of interaction-free measurement" [20]. And by making the atom remote, nonlocal action occurs without observable matter or energy exchange.

## 6. Weak Measurement Enabling Even the Counterfactual Choices in a related Paradox

Finally, let us focus on the case where both parties found their atoms remaining excited and ground, apparently indicating that they were never entangled, according to the naïve reasoning above. Eq. 2 makes it clear that, prior to these two measurements, the atoms have been *entangled*. This is corroborated by the mere *possibility* that the measurements chosen by Alice and Bob could be magnetic moment rather than excited/ground, in which case Bell's proof would be straightforward. This is a simple manifestation of deterministic unitary quantum evolution.

Yet, counterfactuals, by their non-happening, may provide an excuse for an ardent Copenhagenist to deny nonlocality at least to this group of "no-exchange" outcomes.

Can there be a more straightforward argument that even such "no photon exchange" is nonlocally formed?

We next employ weak measurements [10-12] to answer this question. Take all the cases where Alice finds her atom still excited. This time, before this measurement, she performs weak measurements on her atom's magnetic moments $\mu_x$. Not concerned with locality issues this time, Bob, informed about Alice's choice, chooses to make a *strong* measurement of the same magnetic moment.

For a sufficiently large ensemble, when summing up and slicing [11,12] Alice's weak results per Bob's strong ones, a significant correlation appears: The weak outcomes of A's $\mu_x$=+1/-1 correspond to the strong outcomes of B's $\mu_x$=-1/+1, yet Alice's final strong measurement indicates that these correlations came with no photon exchange! For further analysis see Appendix A.

We can take even a step further (see Fig. 2). Assume that the two atoms are ground, and a single photon is emitted towards both by a beam-splitter as in [21], but with a difference in the time of arrival. That is, atom *A* is situated much closer than *B* to the photon's source. When Alice and Bob perform the same measurements as above slightly after the expected arrival time to *A*, they encounter an apparent surprise: Alice's *strong* energy measurement may confirm that the photon has reached her atom, thus never arriving to Bob's atom. However, she performs in addition an earlier, *weak* measurement of her atom's magnetic moment prior to the strong measurement. Bob, on his side, has strongly measured his atom, also for magnetic moment. Weak, nonlocal correlations between the atoms persist, as Alice can check on her side. Therefore, although Alice's excited atomic state implies that the photon has solely interacted with it, entanglement between the two atoms is nevertheless (weakly) detected.[2] In Appendix A we show that this could arise as a disturbance caused by the weak measurement.

---

[2]This is a matter introduced and extensively discussed in [12].

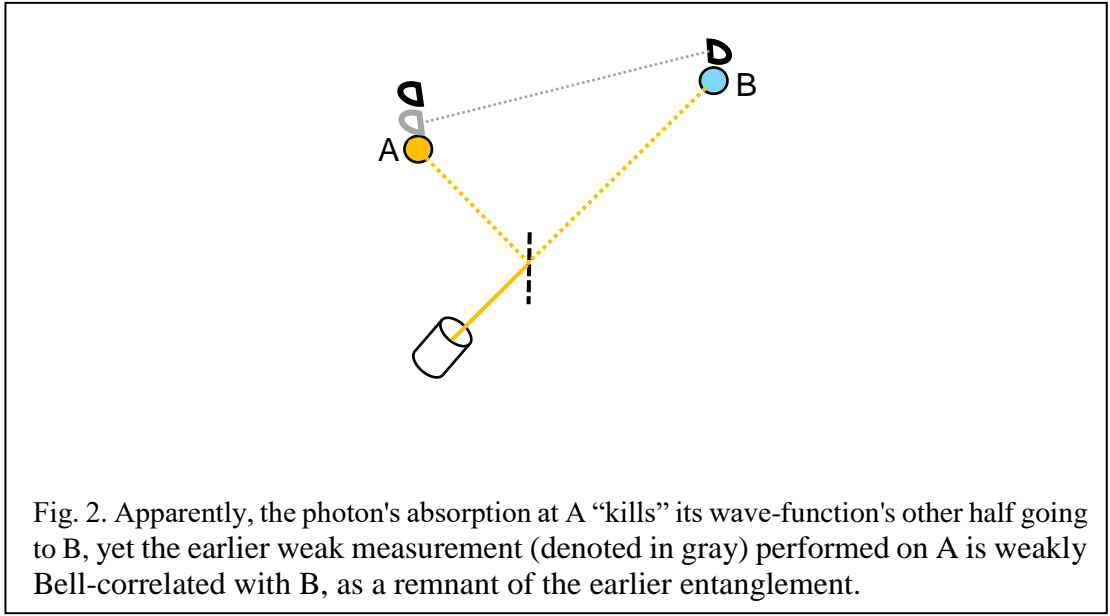

Fig. 2. Apparently, the photon's absorption at A "kills" its wave-function's other half going to B, yet the earlier weak measurement (denoted in gray) performed on A is weakly Bell-correlated with B, as a remnant of the earlier entanglement.

### 7. The Effect in Context

Before concluding, it is worth discussing the Quantum Liar experiment within the broader context of several related quantum mechanical effects, such that they can mutually shed new light on each other. First, to assess the significance of our experiment, let us compare it to its prototype. The standard EPR-Bell experiment involves measurements along three spin directions, out of which Alice and Bob choose one for each particle. These measurements amount to asking the particle "Is your spin up along this direction?" for which the answer may be "yes" or "no."

In the present version, one of the questions is replaced. As entanglement is created by the *possibility* of a photon exchange between the atoms, a measurement set to reveal whether this exchange has occurred metaphorically amounts to asking the atom "Were you entangled with the other atom?" – for which the answer, again, may well imply "no." This paradox renders nonlocality the smallest concern, for in the standard EPR it is taken for granted that the two particles *were* initially parts of the source atom, then *were emitted* towards Alice and Bob, and finally *were absorbed* by their detectors. Nothing of these is granted in the present version, where a photon, naively assumed to be the entanglement's only possible currier, appears to have never been emitted. This assumption turns out to be misleading as in quantum mechanics *possibility* itself gives rise to entanglement. What our analysis shows is that an objective physical effect, namely Oblivion [3], is at the root of this erroneous appearance of no entanglement.

The initial setup of atoms entangled by photon exchange resembles a problem first proposed by Fermi [22] and intensively analyzed later, *e.g.* in [23-26]. However, the rest of that experiment was different, focusing on quantum logic and nonlocality rather than the on the problem of causality, which invoked Fermi's interest in this setup. In this sense, the present analysis may shed new light on that well-studied problem.

Also relevant is "entanglement harvesting" [27,28] allowing two distant atoms to get entangled through their interaction with the electromagnetic vacuum even when spacelike separated. Similarly, "quantum collect calling" allows information transfer without exchange of photons [29,30]. Our experiment, however, is much simpler and more fundamental, not requiring the quantum vacuum as a resource, and moreover persisting regardless of the distance between the atoms.

The present twist in the EPR setting was in fact potentially realizable already by an earlier, major advance employing the EPR-Bell setting, made when Hardy [21] addressed the fundamental question of single-particle nonlocality. Apparently, Bell's proof is not applicable for one particle, because it can be detected by *either* Alice *or* Bob. But Hardy has added an additional measurement, orthogonal to that of the photon's position, with the aid of two atoms on the photon's two paths, thereby creating a full-blown EPR situation.

The setting has been refined in [31] and realized in [32]. The proof therefore holds even when the photon is measured without the mediation of the atoms. Consider, then, the case where Alice chooses a position measurement, detecting the photon on her side. Here too, by Bell's theorem, it is also Bob's remote choice of variable to be measured which has determined this outcome – even when Bob seems to have measured nothing! IFM, then, is thus shown by Bell's theorem to be a nonlocal effect, just like the spin measurement in the standard EPR setting. Other facets of the Liar paradox similarly follow.

The Hong-Ou-Mandel interference [33] can be viewed as a time-reversed version of Hardy's experiment. Two distant low-intensity sources emit single photons towards the same detector within a setting that allows no "which source" information. This gives rise to an interference pattern in the photon's detection place. Elitzur, Dolev and Zeilinger [34,35] have replaced the two sources with two excited atoms (see also the discussions in [36,37]) with a single photon detected from either atom. Under the source-uncertainty in the original HOM, the two atoms become entangled as in Eq. 2, giving an earlier version of the Quantum Liar Paradox.

All these experiments share a common trait: A delicate quantum state is formed, which then undergoes a so-called "quantum oblivion" in the form of a consecutive state that gives the impression that the former state has never occurred [3-5]. A causal gap then emerges, such as a "no click" triggering a distant detector's actual click [1], as well as many other well-known quantum effects stemming from the same dynamics [3].

## 8. Discussion

The Quantum Liar experiment seems paradoxical by its apparent disregard for classical logic: The photon exchange, naively required to facilitate entanglement, seems to have *occurred or not occurred* in accordance with this entanglement. Within quantum theory, however, this classical reasoning has to be changed once it is realized that IFM [1], so far applied only to *spatial* uncertainty, equally applies to the *temporal* one. In the former case, the detector's silence means "The particle is not here, hence it must be in the other possible location." In the present case, with uncertainty plaguing the emission's *timing*, the non-click means "The particle has not been emitted now, hence it must be/have been emitted later/earlier." In both cases, the uncertainty principle takes its toll with an observable effect.

Seeking an appropriate framework to interpret these effects, we might mention that there is nothing in the thought experiments discussed above that is not comprehensible within completions of quantum mechanics such as de Broglie-Bohm theory or the real ensemble formulation [38]. In these formulations, the wave functional is real as in particular are all of its branches. These interfere and entangle with each other, as they satisfy the Schrödinger equation. There are additional elements which are real, and obviate the need for a separate dynamics for measurement; In de Broglie-Bohm this is the particle's position, and in [38] it is the beables of the members of the ensemble. From the point of view of de Broglie-Bohm the wavefunction in the experiment in Fig. 1 has the entangled form of Eq. 2, consisting of the two branches indicated. The actual atoms are either in their excited states or ground states, and the propensity to transition is guided by the entangled wavefunction.

The quantum temporal anomalies involved in this experiment may encourage the use of a time-symmetric framework [39-49], long familiar from Wheeler's "delayed-choice" paradox [50] and made more acute in recent settings like the "too-late choice" experiment [51]. The main reason for that as explained below, is the light shed on the above paradoxes by the *combination* of past and future boundary conditions. While each of them *alone* could give the wrong impression with regard to the interaction that actually took place, together they complement each other, offering a richer account of quantum reality.

A fruitful time-symmetric framework is offered, for instance, by the Two-State-Vector-Formalism (TSVF), where the description of a quantum system is formed by two wave-functions proceeding along both time directions. There are cases where pre- and post-selection seem to give contradictory results. The contradiction, however, is only apparent. Within the TSVF, the information provided by the two boundary conditions about the quantum values prevailing between them is *complementary* [52-54]. Using this formalism and the time-dependent definition of weak values:

$$A_w(t) = \frac{\langle \Phi | U^\dagger(t-t_f) A U(t-t_i) | \Psi \rangle}{\langle \Phi | U^\dagger(t-t_f) U(t-t_i) | \Psi \rangle}, \tag{6}$$

it becomes clear that the time-evolved pre- and post-selection states co-exist at all intermediate instances. Thus, in our Gedanken experiment, weak measurements performed by Alice and Bob reveal the combination of an entangled state given by Eq. 2 (rather than a product $|e\rangle|g\rangle$), and a product state evolving backwards from the future. Together, both measurements give rise to nonlocal correlations, suggesting that Bob's atom has subtly changed despite the absence of observable photon exchange.

These nonlocal correlations can be also understood in terms of the "Cheshire Cat" effect [55]: A particle may take one path while its spin is weakly measured along a different one. An analogous "catless" bare smile (*i.e.* magnetic moment with no energy) may be carried in the form of the born/unborn photon for the present experiment.

This time-symmetric formulation of the paradox may gain further insights when examined within the recently formulated time-symmetric Heisenberg framework [53,54]. Within this formulation, Alice's atom has a deterministic operator with regard to its energy (*i.e.*, when projecting on the atom's energy she finds an excited state with certainty), yet it also has a nonlocal deterministic operator sensitive to the relative phase between $|e\rangle_A|g\rangle_B$ and $|g\rangle_A|e\rangle_B$. It is this operator which generalizes the single-particle notion of modular momentum [56], accounting for the nonlocal correlations with Bob's atom. In other words, the initial (entangled) state of the system suggests that the operator $\sigma_y^A \sigma_x^B$ is a deterministic operator, while the final strong measurements disentangle the state, giving rise to $\sigma_z^A$ and $\sigma_x^B$ as the set of deterministic operators. This kind of complementarity between future and past is discussed in detail in [52-54].

Finally, the fact that the laboratory protocol for the effect's demonstration is already in wide use for practical proposes, such as quantum transmission [9], makes the experiment very feasible. Once an EPR pair is prepared by a possible photon exchange between the two atoms as described above, the now-entangled atoms should undergo Bell tests, selecting the cases where one or two of the atoms underwent energy measurement and found to be in the same state as its preparation, namely ground or excited state. Such cases give the impression that the atom has "remained" in its earlier state, hence no photon seems to have ever been exchanged, yet – and here lies the surprise – Bell's theorem proves that this impression of "remaining" is misleading: The apparent "no entanglement" is a direct consequence of this very entanglement.

While the Quantum Liar experiment appears paradoxical, time-symmetric approaches make it more natural. For example, an analysis performed using the TSVF approach within quantum mechanics suggested that the born/unborn photon has unique physical properties [57,58]. Moreover, recent experiments [59,60] and thought experiments [61-63] employ strong rather than weak measurements for analyzing new phenomena. A subsequent work [18], based on Davies *et al.* [57,58], examines through the analysis of weak values the evolution between two strong "no-emission" measurements: the wave-function is first weakly radiated and then weakly "drawn back" to its still-excited atom. Such combinations of strong and weak measurement offer many further opportunities for exploring the horizons of quantum reality.

**Acknowledgements**

It is a pleasure to thank Sandu Popescu for helpful comments and discussions. We also wish to thank three referees for assisting us to improve the paper. This research was supported in part by Perimeter Institute for Theoretical Physics. Research at Perimeter Institute is supported by the Government of Canada through the Department of

Innovation, Science and Economic Development and by the Province of Ontario through the Ministry of Research and Innovation. A.C.E. wishes to especially thank Perimeter Institute's administrative and bistro staff for their precious help. Y.A. acknowledges support from the Israel Science Foundation Grant No. 1311/14 and ICORE Excellence Center "Circle of Light". E.C. was supported by ERC AdG NLST.

**Appendix A – Further analysis in terms of weak measurements**

As an additional elaboration on the analysis of the paradox using weak measurements, we shall now quantify the interplay between the information gain through weak measurements and their disturbance to the measured state. For doing so, we shall use the mapping of the excited/ground states to the up/down eigenstate of a Pauli-$z$ matrix, and perform the weak measurement in the $x$ basis. Let the weak measurement be described as usual via the von Neumann Hamiltonian:

$$H_{int}(t) = \frac{\lambda}{\sqrt{N}} g(t)\sigma_x p, \qquad (7)$$

where $N$ is the number of measured atoms in Alice's ensemble, and the momentum $p$ is the canonical conjugate of $q$, representing the position of the measuring pointer. The coupling $g(t)$ differs from zero only during the measurement interval $0 \leq t \leq T$ and normalized according to

$$\int_0^T g(t)dt = 1. \qquad (8)$$

Let the initial wavefunction of the system be:

$$\psi = \exp(-q^2)|\downarrow\rangle, \qquad (9)$$

that is, Alice's atom is ground and the pointer is described by a wide Gaussian in comparison to the measurement weakness: $\frac{\lambda}{\sqrt{N}} \ll \frac{\hbar}{\Delta p}$, but we also require that $\lambda \gg \frac{\hbar}{\Delta p}$. These conditions suggest that a single weak measurement provides a negligible amount of information, but when performing the weak measurement repeatedly over a large ensemble, the average translation of the point grows like $\lambda\sqrt{N}$, while the uncertainty grows like $\frac{\hbar}{\Delta p}\sqrt{N}$ (as known for normal random variables) [11,12]. Hence, the averaged translation of the pointer when measuring the pre- and post-selected ensemble would be significant.

But how much disturbance was induced by each measurement? Let us examine the time evolution of the system (assuming $\hbar = 1$ for simplicity):

$$\exp\left(-\frac{i\lambda}{\sqrt{N}}\int_0^T g(t)\sigma_x P_d\right)\exp(-q^2)|\downarrow\rangle = \frac{1}{\sqrt{2}}\exp\left(-\frac{i\lambda}{\sqrt{N}}\sigma_x P_d\right)\exp(-q^2)(|\rightarrow\rangle - |\leftarrow\rangle) =$$

$$\frac{1}{\sqrt{2}}\left\{\exp\left(-(q-\lambda/\sqrt{N})^2\right)|\rightarrow\rangle - \exp\left(-(q+\lambda/\sqrt{N})^2\right)|\leftarrow\rangle\right\} \simeq$$

$$\simeq \frac{1}{\sqrt{2}}\left\{\left[1-(q-\lambda/\sqrt{N})^2\right]|\rightarrow\rangle - \left[1-(q+\lambda/\sqrt{N})^2\right]|\leftarrow\rangle\right\} =$$

$$= \frac{1}{\sqrt{2}}\left\{\left[1-q^2-\frac{\lambda^2}{N}+\frac{2\lambda q}{\sqrt{N}}\right]|\rightarrow\rangle - \left[1-q^2-\frac{\lambda^2}{N}-\frac{2\lambda q}{\sqrt{N}}\right]|\leftarrow\rangle\right\} =$$

$$= \left(1-q^2-\frac{\lambda^2}{N}\right)|\downarrow\rangle + \frac{2\lambda q}{\sqrt{N}}|\uparrow\rangle \qquad (10)$$

Then the pointer's shift is read and a translation of some *q'* arises in each run. When repeated over the large ensemble, this allows to infer the weak value, but also suggests the negligible flip chance of $\frac{4\lambda^2 q'^2}{N}$, i.e. a small disturbance to the measured state. As mentioned above, and as must be the case in order to maintain the uncertainty principle, this is also the amount of information provided by a single weak measurement of the ground atom in our experiment. So is it possible to explain this way the nonlocal correlations between Alice and Bob which are described in a simple way using the TSVF? Yes, otherwise the latter would not have been equivalent to quantum mechanics. But is such an explanation plausible? This is a somewhat philosophical question, going beyond the scope of this paper, but first, it is possible to make the measurement strength smaller and smaller while correspondingly increasing the ensemble's size. The effect would still persist even though the amount of disturbance in each single experiment is very small [11,12]. Another reason (discussed in [11,12]) for the TSVF being heuristically simple is the apparent implausibility of assigning a property to an ensemble of $N \gg 1$ atoms based on a very few atoms $4\lambda^2 q'^2$ that went from ground to excited. We find it more natural to assign this property to each of the atoms individually (this was recently supported by [64]). In a recent series of papers [52-54], we have further advocated this view, based on a realistic and deterministic account of QM relying on the combination of two boundary conditions. For a related discussion on this topic see [65,66].

In any case, weak measurement is used here as a complementary tool for studying the paradox which is just as acute with projective (strong) measurements. As discussed in the main text, there seems to be more than one way to consistently interpret the results.